\documentclass{mn2e}
\usepackage{epsfig}
\def\msun{{\rm M_{\odot}}}

\def\mo{{\dot M_{\rm out}}}

\def\ngc{{NGC 4051}}

\def\msun{{\rm M_{\odot}}}

\def\mo{{\dot M_{\rm out}}}

\def\xmm{{\it XMM-Newton}}
\def\chandra{{\it Chandra}}

\def\et{{et al.\ }}

\def\xte{{\it RXTE}}

\newcommand{\ls}{\mathrel{\hbox{\rlap{\hbox{\lower4pt\hbox{$\sim$}}}\hbox{$<$}}}}
\newcommand{\gs}{\mathrel{\hbox{\rlap{\hbox{\lower4pt\hbox{$\sim$}}}\hbox{$>$}}}}

\def\Msun{\hbox{$\rm ~M_{\odot}$}}

\def\H0{{\rm ~km~s^{-1}~Mpc^{-1}}}

\def\msun{M_{\rm \odot}}

\def\et{{et al.}}

\title[X-ray spectrum of \ngc]
{An extended \xmm\ observation of the Seyfert galaxy \ngc.  
I. Evidence for a shocked outflow}
\author[K.A.Pounds \et]
        {K.A.Pounds,
	S.Vaughan 
	\\
Department of Physics and Astronomy, University of Leicester,
Leicester, LE1 7RH, UK\\}

\date{Accepted ; Submitted }
\pagerange{\pageref{firstpage}--\pageref{lastpage}}
\pubyear{2010}
\begin{document}
\maketitle
\label{firstpage}

\begin{abstract}    An extended \xmm\ observation of the Seyfert 1 galaxy \ngc\ has revealed a rich absorption line spectrum indicating the presence of a
photoionised outflow with a wide range of velocities and ionisation parameter. At low continuum fluxes an emission line spectrum is well defined with both
narrow and broad emission components of several abundant metal ions. The absorption line velocity structure and a broad  correlation of velocity with
ionisation parameter are consistent with an outflow scenario where a  highly ionised, high velocity wind, perhaps launched during intermittent
super-Eddington accretion, runs into the interstellar medium or previous ejecta, losing much of its kinetic energy in the resultant strong  shock. We
explore the possibility that a quasi-constant soft X-ray component may be evidence of this post-shock cooling. This revised view of AGN outflows is
consistent with multiple minor Eddington accretion episodes creating a momentum-driven feedback linking black hole and host galaxy growth.    
\end{abstract}


\section{Introduction}

\ngc\ is a bright, narrow line Seyfert 1 galaxy in the Ursa Major cluster, lying at a Tully-Fisher distance of  15.2 Mpc (Russell 2003), with  a
heliocentric velocity of 753 km s$^{-1}$ (Verheijen 2001). The rapid and large  amplitude variability (Lawrence \et\ 1985,1987) was strong early evidence
that the powerful X-radiation found to be a common property of Seyfert galaxies by Ariel 5 (Cooke \et\ 1976, Elvis \et\ 1978) and Uhuru (Tananbaum \et\
1978) observations resulted from accretion onto a central supermassive black hole. Subsequent monitoring with \xte\ revealed a different aspect of \ngc,
where it occasionally lapses into extended periods of low and quasi-constant X-ray emission (Lamer \et\ 2003). More recently, high resolution X-ray
spectra provided by \chandra\ and \xmm\ have shown that \ngc\ also exhibits a strong ionised outflow.

An early \chandra\ HETG observation resolved two X-ray absorption line systems, with outflowing velocities of photoionised gas of $\sim$600 and $\sim$2300
km s$^{-1}$, while contemporaneous HST spectra of CIV, NV and SiIV found several absorption systems with velocities from $\sim$30 km s$^{-1}$ to $\sim$650
km s$^{-1}$ (Collinge \et 2001). Of particular interest in the context of the present analysis, where we find a clear correlation of velocity and
ionisation parameter, the higher velocity X-ray component had no counterpart in the UV spectra. The \chandra\ data also showed an unresolved Fe K emission
line at $\sim$6.41 keV (FWHM $\leq$2800 km s$^{-1}$).

Observations of \ngc\ with \xmm\ in 2001 and 2002 coincided with periods of relatively high and low X-ray flux, offering an opportunity to further explore
the complexity of its X-ray spectrum. Pounds \et\ (2004), hereafter Po04, found the hard X-ray band to be dominated by reflection from cold matter, which
could also explain a non-varying, narrow Fe K fluorescent line. A soft X-ray narrow emission line spectrum evident at low continuum fluxes, with observed
wavelengths consistent with the \ngc\ rest frame, indicated an extended ionised emission region, while a dominant absorption line spectrum in the high
flux RGS observation of 2001 showed an ionised outflow  with a line of sight velocity of $\sim$500 km s$^{-1}$. An Fe K absorption line in the  EPIC
spectrum indicated the presence of a more highly ionised outflow component with a velocity some 10 or 30 times higher, depending on the ionisation
state being FeXXVI or FeXXV.

Most recently, Steenbrugge \et\ (2009) have reported the analysis of \chandra\ LETG observations in 2001 and 2003, finding evidence for a more complex
ionised outflow, with low ionisation components (log$\xi$$\sim$0.1 and log$\xi$$\sim$0.5-0.9) outflowing at v$\sim$200-330 km s$^{-1}$, a more highly
ionised  component of log$\xi$$\sim$2 at v$\sim$600 km s$^{-1}$ and a high ionisation component (log$\xi$$\sim$3) at v$\sim$4600 km s$^{-1}$.  Those
authors find no evidence for recombination in 3 of the 4 ionisation components up to 20 ks after a sudden, factor-of-5 drop in  X-ray flux, indicating 
their location at a radial distance $\ga$$7\times 10^{16}$ cm for the higher velocity gas and $\ga$$9\times 10^{17}$ cm for the lowest velocity absorption
in OV. We note that these minimum radii conflict with those from a similar time variability analysis of Krongold \et\ (2007), and have the incidental
bonus of removing the conceptual problem in Krongold \et\ where the local escape velocity exceeded the measured outflow value.

Evidence for much higher velocity X-ray outflows (Chartas \et\ 2002, Pounds \et\  2003, 2006; Reeves \et\ 2003, Cappi 2006, Tombesi \et\ 2010) has been
confined to the very highly ionised matter (log$\xi$$\sim$3.5--4) most readily detected in the Fe K band. The high velocities and high column densities in
these Fe K band observations are thought to offer the best prospect that such energetic flows  represent a significant feedback mechanism to constrain the
continued growth of a black hole and star formation in the host galaxy, although to date only in the case of the  bright QSO PG1211+143 has the requisite
wide angle flow and large covering factor been directly observed (Pounds and Reeves 2009).  In the case of \ngc\,in contrast, Krongold \et\ (2007)
concluded that the low velocity gas has a small covering factor and hence involves a relatively insignificant mass and energy rate. While Steenbrugge \et\
(2009) do not confirm the high density and small radial distance of the ionised outflow, they also  derive a low mass and energy rate from the soft X-ray
spectrum of \ngc. 

In this paper we report an analysis of a new \xmm\ study of \ngc\ with substantially greater sensitivity than hitherto, finding a rich absorption line
spectrum revealing a velocity-structured outflow covering a wide range in velocity and ionisation parameter. We re-examine the suggestion in Po04 that
both high and low velocity absorption lines arise at different stages of a mass-conserved outflow. Our new assessment considers the new data in the
context of a high velocity, highly ionised wind being slowed by interaction with the interstellar medium, losing much of its mechanical energy in the
resulting shock, while retaining momentum to push low ionisation gas out into the host galaxy (King 2010). We note that the same ideas would
apply to an outflow colliding with earlier, slower moving ejecta.

The wider importance of such shocked outflows, perhaps launched during an Eddington accretion episode (King and Pounds 2003),  lies in the possibility that
the accumulated thrust from multiple episodes - rather than the outflow energy - would eventually drive gas from the bulge, thereby limiting further star
formation and black hole growth. Such a momentum-driven feedback mechanism has been shown by King (2003,2005) to reproduce the observed  correlation of
black hole and galaxy masses (e.g. Ferrarese and Merritt 2000, Gebhardt \et\ 2000, Haring and Rix 2004)). 

\section{Observations and data analysis}

\ngc\ was observed by \xmm\ on 15 orbits between 2009 May 3 and June 15. Here we use mainly the soft X-ray data from the Reflection Grating
Spectrometer/RGS (den Herder \et\ 2001), while also checking for evidence of a more highly ionised outflow in the EPIC pn (Str\"{u}der \et 2001) and MOS
(Turner \et\ 2001) spectra. Excluding high background data near the end of each orbit the net exposures available for spectral fitting were typically 40
ks per orbit, yielding an overall exposure of $\sim$600 ks (pn) and $\sim$1.2 Ms (combined MOS and RGS), factors of $\sim$6  and $\sim$12 higher than in
the \xmm\ observations of 2001 and 2002. In the present paper we use spectra integrated over whole orbits, and note the mean orbit flux levels cover a
similar range to the `bright' and `faint' observations of 2001 and 2002. Full details on the timing, flux levels and X-ray light curves for each revolution
are included in an accompanying paper on the `Rapid X-ray variability of \ngc' by Vaughan \et\ (2010).

\begin{figure*}                                                          
\centering                                                              
\includegraphics[width=6.8cm, angle=270]{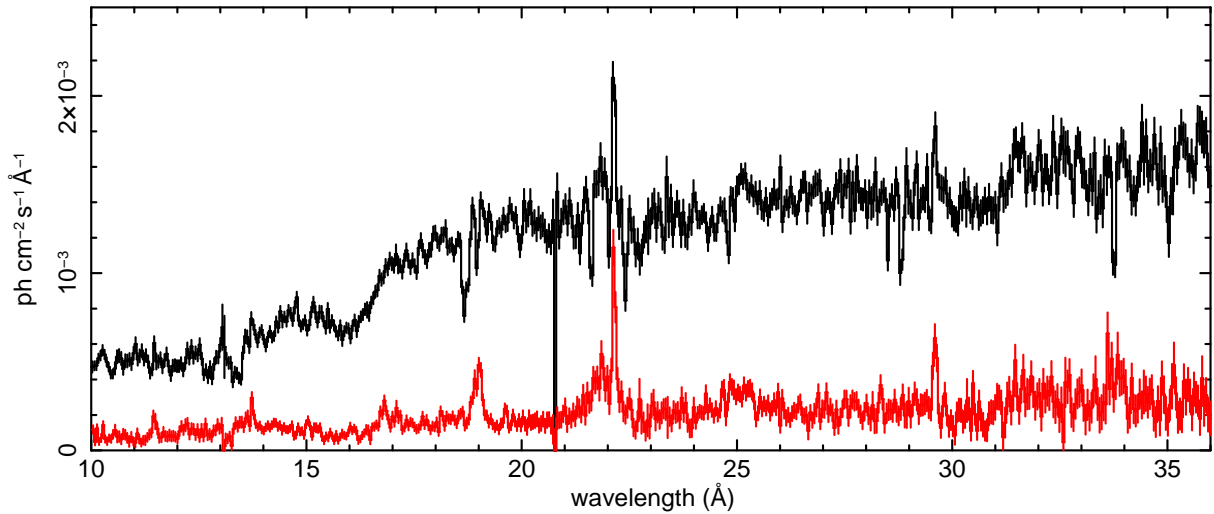}                     
\caption                                                                
{Fluxed RGS spectra of \ngc\ summed over 4 relatively bright and 3 faint continuum levels illustrating the change from a complex
absorption line spectrum to a dominant emission line spectrum as the continuum flux falls}      
\end{figure*}

Figure 1 contrasts the RGS spectrum at representative high (revs 1727-1730) and low (revs 1725,36,39) flux levels during the 2009 observation. The 
distinction found previously (Po04)  is again evident, with a dominant absorption line spectrum seen against the stronger continuum when bright, and an
emission line spectrum emerging more clearly when the continuum is faint.  The most obvious absorption lines are associated with H- and He-like ions of C,
N, O and Ne, and multiple velocity components are seen for several resonance transitions. The emission spectrum is also better defined than in any earlier
observation of \ngc, benefiting from the long exposures of the 2009 observation. 

Reference to figure 1 also shows a broad trough in the RGS spectrum between $\sim$15.5-16.8 \AA, present in both high and low flux data, which we attribute
to a UTA from Fe-L absorption in low ionisation matter. Inclusion of this UTA casts doubt on the reality of a relativistic emission line of OVIII
Lyman-$\alpha$, previously reported for \ngc\ by Ogle \et\ (2004) and Steenbrugge \et\ (2009).

In analysing the 2001 \xmm\ absorption spectrum of \ngc, Po04 first subtracted line emission determined in the low flux spectrum of 2002.  However, in the
2009 \xmm\ data  the low flux spectrum still shows residual absorption. We therefore adopted a different  approach,  by first modelling the emission 
spectrum in the low flux data and retaining the measured emission
parameters in fitting the high flux absorption spectrum. 

A representative `highflux' spectrum was obtained by integrating RGS 1 and RGS 2 data from the 4 orbital revolutions 1727-1730, while broad resonance
emission lines and radiative  recombination continua (RRC) were determined from a representative `lowflux' spectrum (revs 1725, 1736 and 1739).  The 
`highflux' continuum was first modelled by a power law ($\Gamma$$\sim$3.5), with broad emission features added as observed in the low flux
spectrum. Individual absorption lines were then fitted with negative  Gaussians, with wavelength, line width and amplitude
as free parameters. 

We discuss the emission line spectrum of \ngc\ in more detail in a separate paper (Pounds and Vaughan 2011; hereafter Paper II). For the present purposes,
in order to quantify the absorption line spectrum in
the highflux state, we simply represent the broad line emission component of each principal resonance line with a positive Gaussian, with parameters
determined in the corresponding lowflux spectrum. 

\section{A complex absorption line spectrum}

\subsection{OVIII Lyman-$\alpha$}

Figure 2 (top panel) illustrates the lowflux spectrum in the region of the OVIII Lyman-$\alpha$ line (rest wavelength 18.97\AA).  A broad emission line 
is clearly resolved, with
substantial absorption near the line core. Fitting a positive Gaussian to the broad emission component finds a rest wavelength of 18.96$\pm$ 0.02 \AA, 
1$\sigma$ line width of 90$\pm$ 9 m\AA\ and amplitude of $3.7\times10^{-4}$ photons cm$^{-2}$ s$^{-1}$ \AA$^{-1}$. The median wavelength indicates a
velocity blueshift of $\sim$750$\pm$200 km s$^{-1}$ while, allowing for the RGS 1 resolution ($\sigma$=28 m\AA\ at 19\AA), the line width corresponds to
$\sim$3100$\pm$400 km s$^{-1}$ FWHM, if interpreted purely as velocity broadening. 

Figure 2 (lower panel) shows the same spectral region near the OVIII Lyman-$\alpha$ line, with a complex of absorption lines superimposed on the 
continuum plus broad
line emission.  A sweep across the 18--20 \AA\ waveband with a negative Gaussian of width initially set to the RGS resolution, found 3 strong and 3 weaker
absorptions lines, located at wavelengths labelled (a) - (f) in the figure. However, it is apparent that the absorption structure is only approximately
matched by Gaussian lines, with the broad opacity between $\sim$18.5-18.8 \AA\ suggesting a spread of velocities.

\begin{figure}
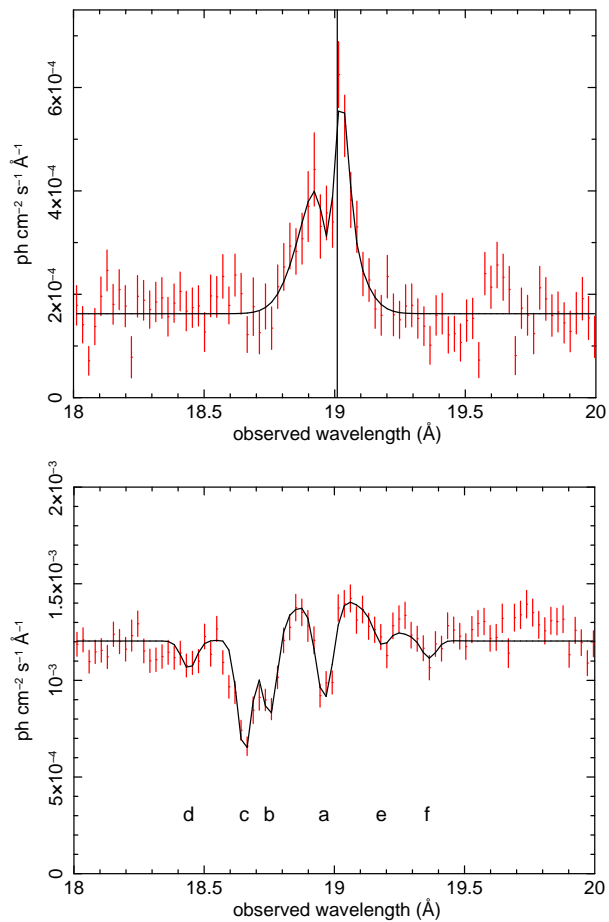
                                                          
\centering                                                              
\includegraphics[width=5.8cm, angle=270]{O8_em.ps}                     
\centering                                                              
\includegraphics[width=6.3cm, angle=270]{O8_abs.ps}  
\caption                                                                
{(top) A broad emission component of OVIII Lyman-$\alpha$ (rest wavelength shown by the vertical line) in the lowflux spectrum of \ngc\ together 
with narrow emission and absorption components.
(lower) Gaussian fitting to absorption lines observed in the highflux continuum spectrum plus broad emission line.  Gaussians labelled a, b, c and d are 
attributed to
outflow components of increasing velocity in OVIII Lyman-$\alpha$. Gaussians e and f are identified with low velocity absorption in OVI}
\end{figure}

Identifying the absorption components (a) - (d) with OVIII Lyman-$\alpha$  yields line-of-sight velocities of -780$\pm$70 km s$^{-1}$, -4100$\pm$85 km
s$^{-1}$, -5600$\pm$60 km s$^{-1}$ and -9000$\pm$300 km s$^{-1}$.     We note that the probable blend of component (c) with the OVII 1s-3p line (rest
wavelength 18.63 \AA) renders the parameters of that component uncertain.   However, velocity  components a, b and c are all independently supported by their
detection in OVIII Lyman-$\beta$ (Table 2).  

The weaker absorption lines labelled (e) and (f) in figure 2 are most likely due inner shell transitions in OVI (rest wavelengths 19.18 \AA\  and 19.34 \AA;
Holczer \et\ 2009), and we note a further possible satellite line in OV (19.97 \AA), all 3 low ionisation lines yielding a low outflow velocity. 

In summary, we find evidence for 3, or possibly 4 blue-shifted velocities in OVIII Lyman-$\alpha$, with a clear separation between the component at $\sim$780
km s$^{-1}$ and components at $\sim$4100 and $\sim$5600 km s$^{-1}$.

\subsection{OVII 1s-2p triplet}

The lowflux spectrum in the region of the OVII 1s-2p triplet also shows a strong and broad emission feature (figure 3, top panel), together with a  narrow
emission line readily identified with the forbidden transition (rest wavelength 22.101 \AA). Fitting a Gaussian to the narrow line  yields a wavelength 
(adjusted for the redshift of \ngc) of 22.092$\pm$ 0.003 \AA, line width $\sigma$=28$\pm$ 3 m\AA\ and amplitude $9.5\times10^{-4}$  photons cm$^{-2}$ s$^{-1}$
\AA$^{-1}$. We find the  OVII forbidden line to be marginally blue-shifted, with a velocity of -125$\pm$40 km s$^{-1}$,  and to be unresolved by RGS1, the 1
$\sigma$ upper limit  corresponding to a velocity width of $\sim$250 km s$^{-1}$ FWHM. 

A single Gaussian fit to the broad emission from the OVII 1s-2p triplet finds a rest wavelength of 21.8$\pm$0.1 \AA, line width $\sigma$=305$\pm$35 m\AA\  
and amplitude $2.2\times10^{-4}$ photons cm$^{-2}$ s$^{-1}$ \AA$^{-1}$.  The intrinsic width, corresponding to a velocity width of $\sim$9500 km s$^{-1}$
(FWHM), is clearly much larger than that found for OVIII Lyman-$\alpha$. However, a more detailed analysis of the emission spectrum in Paper II finds broad
emission components for both the OVII resonance (21.602\AA) and intercombination (21.807 \AA) lines, with component line widths comparable to that found here
for OVIII Lyman-$\alpha$. For the present purposes of quantifying the highflux absorption spectrum we retain the single emission profile, shown added to the
best-fit continuum in figure 3 (lower panel).

Six absorption lines are detected in the region of the OVII triplet, all but that near 22.75 \AA\ being of width consistent with the RGS resolution. 
Those labelled (a), (b), (c) are identified as blue-shifted components of the OVII 1s-2p resonance line, the individual blue shifts yielding  line-of-sight 
outflow velocities of -440$\pm$60 km s$^{-1}$, -4080$\pm$140 km s$^{-1}$, and -5810$\pm$120 km s$^{-1}$, respectively. While  these values are consistent
with the 3 stronger components seen in OVIII Lyman-$\alpha$, the high velocity absorption is much weaker in OVII. We note the strong low velocity
component in OVII is probably a blend of core absorption
in the broad emission line (confirmed as self absorption in Paper II) and absorption in the
highflux continuum.  

Deep absorption lines are also identified with inner shell transitions in OVI (d), OV (e) and OIV (f), confirming the presence  of
substantial low ionisation gas in the outflow from \ngc. The deduced velocities for these low ionisation lines are comparable to the lowest
velocity component in OVII. Detailed values are listed in Table 2, apart from line (f),
where a probable  blend of OIV lines at 22.74 \AA\ and 22.77 \AA\ (Holczer \et\ 2009) makes a velocity  measurement uncertain.

\begin{figure}
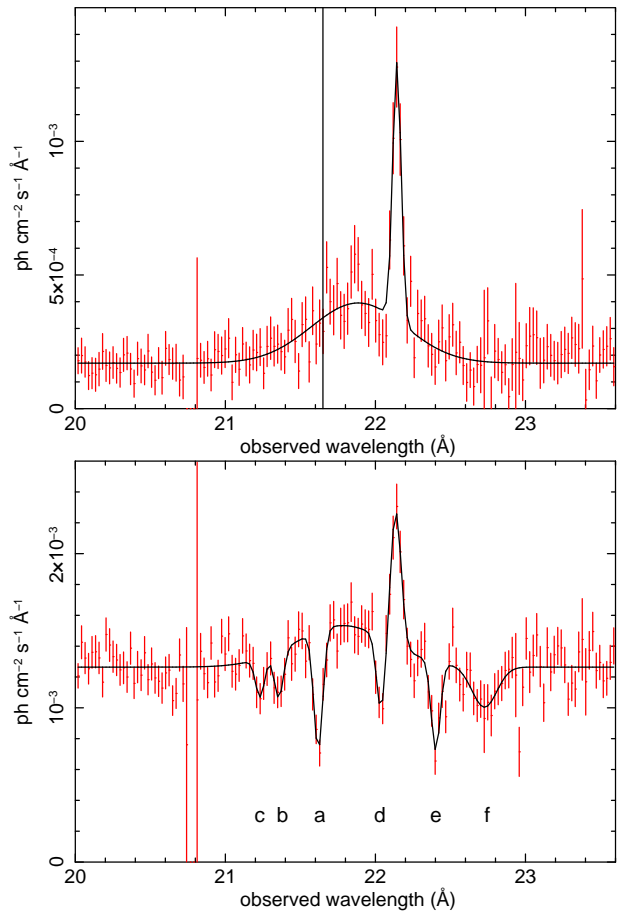
                                                                              
\centering                                                              
\includegraphics[width=6cm, angle=270]{sav3_rev.ps}                     
\centering                                                              
\includegraphics[width=6cm, angle=270]{sav6_rev.ps}                      

\caption                                                                
{(top) A broad emission line in the lowflux spectrum at the OVII triplet is a blend of components from the resonance and
intercombination lines. An unresolved emission component is identified with the forbidden line in the OVII triplet.  
(lower) Gaussian fitting to absorption lines in the highflux spectrum in the wavelength region of the OVII triplet. Those labelled a, b and 
c are identified as blue-shifted components of the OVII 1s-2p  resonance transition (rest wavelength 21.60 \AA), while 
components d, e and f are identified with the low velocity flow in OVI, OV, and OIV, respectively}      
\end{figure}

\subsection{NVI 1s-2p and CVI Lyman-$\beta$}
 
Given the complexity of the emission and absorption in the region of the OVII triplet, it is interesting to examine the RGS data for other He-like ions. That
for CV unfortunately lies beyond the range of the RGS, while the spectrum in the vicinity of the NeIX triplet is complicated by the presence of strong Fe-L
lines. That makes NVI the best candidate for direct comparison with OVII.

Figure 4 shows the highflux data in the region of the NVI triplet. As for OVII, a broad emission line has been added to the power law  continuum as a
baseline. The forbidden line (rest wavelength  29.534 \AA) is again unresolved and marginally blue-shifted. 

The strongest absorption lines, labelled (a)  and (b), are readily identified with the low velocity outflow component seen  in the 1s-2p resonance line  of
NVI (rest wavelength 28.787 \AA) and CVI Lyman-$\beta$ (rest wavelength 28.466 \AA), respectively. The deduced velocities of -325$\pm$50 km s$^{-1}$ and
-260$\pm$40 km s$^{-1}$, the lowest found in the analysis so far, extend the observed link of velocity with ionisation energy.

We note that any higher velocity component  of NVI, corresponding to that observed near 4000 km s$^{-1}$ in OVIII and OVII, would be blended with the
absorption in CVI. Possible interpretations of the absorption lines (c), (d), (e) include lower ionisation stages of N, but the identifications and rest
wavelengths are not sufficiently secure to derive useful flow velocities.     

\begin{figure}                                                          
\centering                                                              
\includegraphics[width=6cm, angle=270]{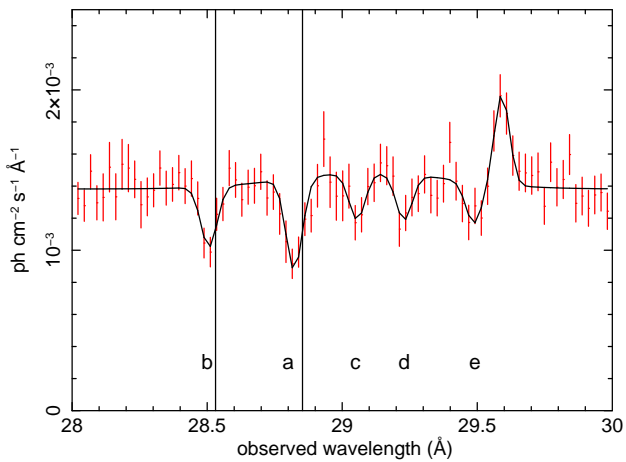}                                          
\caption                                                                
{Absorption lines in the highflux
spectrum reveal low outflow velocity components of the NVI 1s-2p resonance line (a) and CVI Lyman-$\beta$ (b). The vertical lines again 
correspond to the rest wavelengths of these transitions}      
\end{figure}

\subsection{Lyman-$\alpha$ of CVI, NVII and NeX} 

Outflow velocities found in OVIII Lyman-$\alpha$ at $\sim$5600 and $\sim$9000 km s$^{-1}$ are considerably higher than any previously reported from \ngc.
However, the statistical significance of the highest component in OVIII is marginal, underlining  the need to look for evidence of a similar velocity
structure in other H-like resonance  transitions available within the RGS waveband. Those are of CVI, NVII and NeX. In modelling the highflux absorption
spectrum, a shallow emission line of width $\sigma$=90 m\AA\ is added to the best-fit highflux continuum  in each case.

Figure 5 (top panel) shows the RGS spectrum in the waveband 32 - 36 \AA, which covers the region of CVI Lyman-$\alpha$, and the 1s-3p transition of CV. The
same 4 velocity components detected in OVIII Lyman-$\alpha$ may be seen as absorption lines (a), (b), (c) and (d) in CVI Lyman-$\alpha$  (rest wavelength
33.74 \AA), with values here of -490$\pm$35 km s$^{-1}$, -4000$\pm$70 km s$^{-1}$, -5900$\pm$90 km s$^{-1}$, and -8920$\pm$180 km s$^{-1}$. Identifying the
absorption line (e) with a low outflow velocity in CV 1s-3p (rest wavelength 34.97 \AA), of -250$\pm$60 km s$^{-1}$, further extends the trend where the
lowest outflow velocities have the greatest opacity in the lowest ionisation stages. 

Figure 5 (mid panel) shows absorption lines in the spectral band encompassing the NVII Lyman-$\alpha$ line (rest wavelength 24.781 \AA), and the 1s-3p
resonance line
of NVI (24.90 \AA). A Gaussian sweep finds 5 absorption lines. Those labelled (a), (b), (c) and (d), yield outflow velocities of  
-495$\pm$50 km s$^{-1}$, -4000$\pm$80 km s$^{-1}$, -5920$\pm$135 km
s$^{-1}$, and -8600$\pm$120 km s$^{-1}$, consistent with
values seen in OVIII and CVI. For NVII Lyman-$\alpha$, the deduced velocities are .  Component (e) corresponds to an outflow at -325$\pm$50 km s$^{-1}$ 
if correctly attributed to NVI. The positive Gaussian at
$\sim$25.3 \AA\ is an approximate match to the strong RRC of  CVI. 

\begin{figure}
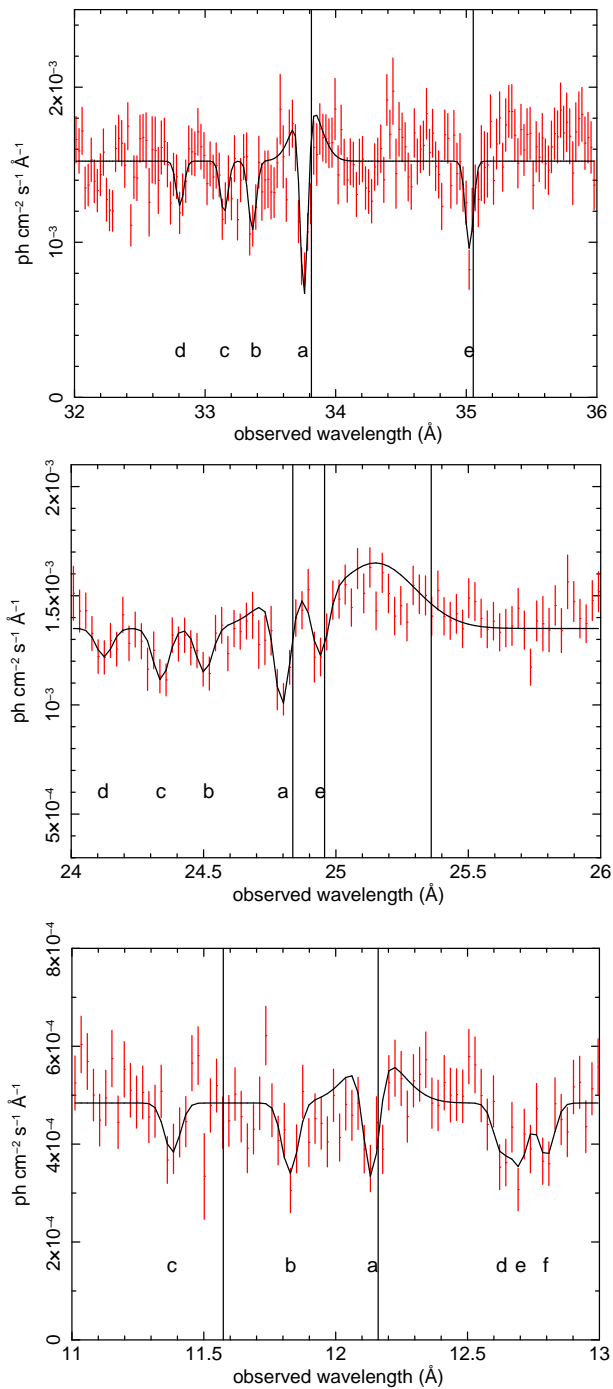
                                                          
\centering                                                              
\includegraphics[width=5.8cm, angle=270]{CVI_abs.ps}
\centering                                                              
\includegraphics[width=6.1cm, angle=270]{NVII_abs.ps}                                          
\centering                                                              
\includegraphics[width=6.38cm, angle=270]{NeX_abs.ps}                                                                                    
\caption                                                                
{(top) Absorption line spectrum in the wavelength region  of CVI Lyman-$\alpha$ and the 1s-3p transition of CV. 
Lines labelled a, b, c and d are identified with velocity components in CVI Lyman-$\alpha$ consistent with
velocities found in  OVIII. Line e is identified with a low velocity outflow in CV. (middle) Absorption lines
in the wavelength region  of NVII Lyman-$\alpha$ and the 1s-3p transition of NVI. Lines labelled a, b, c
and d are identified with velocity components in NVII Lyman-$\alpha$ similar to those found in OVIII and CVI.  
Line e is identified with a low velocity components of NVI 1s-3p. 
(lower) Absorption lines in the
wavelength region  of NeX Lyman-$\alpha$. Lines labelled a and b correspond to the lowest and highest velocity outflow components 
found in OVIII, CVI and NVII. Line c corresponds to an intermediate velocity in NeIX, while lines d, e and f are probably identified with FeXX or FeXIX, 
but the
uncertain identification limits their value as a  velocity indicator}      
\end{figure}

The final Lyman-$\alpha$ line accessible in the RGS waveband is that of NeX, having the highest ionisation potential and shortest wavelength (rest wavelength
12.13 \AA). Figure 5 (lower panel) shows the highflux absorption  spectrum. Although the RGS suffers from decreasing sensitivity in this short wavelength region,
several absorption lines are seen. Line (a) lies just shortward of NeX Lyman-$\alpha$ (rest wavelength 12.13 \AA), and is most probably due to a low
velocity component of -620$\pm$150 km s$^{-1}$. If also attributed to NeX, line (b)  corresponds to a much higher outflow  component at -8230$\pm$130 km
s$^{-1}$, while intermediate velocity components at $\sim$4000 and $\sim$6000 km s$^{-1}$ are not detected in NeX. We suggest the most likely 
identification of line (c) is with NeIX 1s-3p (rest wavelength 11.55 \AA), at a velocity of -5000$\pm$350 km s$^{-1}$ It is interesting that for NeX only the
highest and lowest velocity components of those seen in OVIII, NVII and CVI are detected, a further indication of  the trend of increasing outflow velocity
with  ionisation energy of the parent ion. 

Absorption lines (d), (e) and (f) are probably identified with FeXX or FeXIX, but the uncertain individual transitions makes  their present value
as a velocity indicator not very useful.

In summary, the velocity structure is similar in all 4 Lyman-$\alpha$ lines visible in the RGS spectrum, all showing velocity components up to
$\sim$8000--9000  km s$^{-1}$, in addition to a well-separated low velocity component. The highest velocity of $\sim$8500 km s$^{-1}$ is notably
strongest in the highest energy Ne ion, extending the trend of velocity being correlated with ionisation parameter at least in the high velocity gas.  We
note, in contrast, that the low velocity absorption is seen in all resonance lines across the RGS spectrum. 

In Section 6 we suggest that this
difference, also reflected in XSTAR modelling, is due to a separate origin of a component of the low velocity absorption arising from self absorption 
in  a
limb-brightened shell, rather than in the direct AGN continuum.

\subsection{A velocity- and ionisation-structured outflow}
 
Table 2 brings together all the outflow velocities assigned to a significant absorption component where the parent ion can be reliably identified.  Also
listed in Table 2 is the ionisation parameter at which each ion stage would have a maximum concentration in a photoionised gas illuminated by the AGN
continuum flux (Kallman \et\ 1996). Several conclusions can be drawn from the data in Table 2, which are also shown graphically in figure 8. 

First, several discrete outflow velocities are indicated, each across a range of different ions, with a clear distinction between a `low velocity' group 
($\sim$250-800 km s$^{-1}$) and a higher velocity group, with components at $\sim$4000, $\sim$6000 and $\sim$8500 km s$^{-1}$. Second, there is a strong
correlation of velocity with ionisation parameter, absorption in the lower velocity components being strongest in the lower ionisation lines, and the higher
velocity components being noticeably stronger for the higher ionisation lines. Moreover, only the lowest velocity flow is evident in the lowest ionisation
gas. 

To ascertain whether the trend of increasing outflow velocity with ionisation parameter continues to still higher values, we look  for evidence of absorption
lines in the EPIC data, covering the K-shell energies of the heavier abundant elements Mg, Si, S. Ar and Fe.

\section{Absorption lines in the EPIC data}
\subsection{Intermediate mass ions of Mg to Ar}

Examination of the EPIC spectra reveals several absorption lines, detected in both pn and MOS data, lying close to K-band resonance  transitions
in Mg, Si, S, Ar and Fe. 

Figure 6 shows a section of the highflux (revs1727-30) pn and MOS spectra in the energy range from 1-4 keV.  Although the plotted ratio of data to a simple
power law continuum fit are rather noisy (with deviations near 1.8 keV and 2.3 KeV that are probably due to imperfect modelling of the Si K and Au M edges in
the instrument response function), a consistent set of absorption lines is seen for the 4 abundant elements with K-shell resonance transitions in this energy
range.

Identifying the features, observed in both pn and MOS data near 1.50, 2.05, 2.68 and 3.42 keV, with Lyman-$\alpha$ absorption lines of Mg, Si, S and Ar,
respectively, and taking a weighted mean of Gaussian fits to the MOS and pn data, yields  outflow velocities of -7100$\pm$1500 km s$^{-1}$, -7100$\pm$2000 km
s$^{-1}$, -7700$\pm$1600 km s$^{-1}$ and -9700$\pm$1800 km s$^{-1}$ km s$^{-1}$.  Within the substantially larger uncertainties these values are consistent
with the higher velocity components in the RGS data,  extending their detection over a still wider range of ionisation parameter. 

The EPIC velocities are added to those from the RGS analysis in Table 2.

\begin{figure}
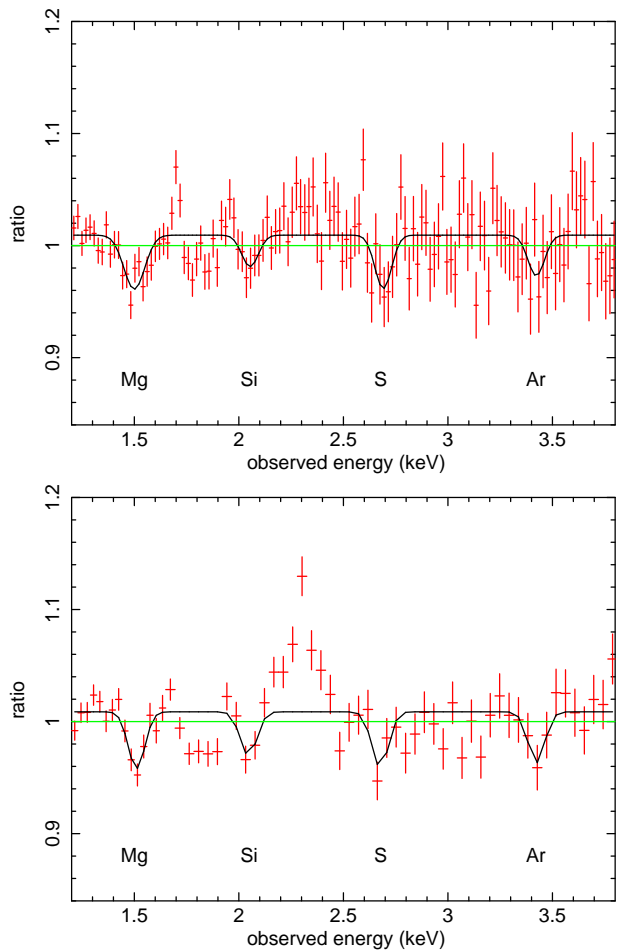
                                                          
\centering                                                              
\includegraphics[width=6.3cm, angle=270]{pn.ps}                     
\centering                                                              
\includegraphics[width=6.3cm, angle=270]{mos.ps}                     
\caption                                                                
{Negative Gaussian fits to (top) the highflux pn data and (lower) the corresponding MOS data, plotted in both cases against  a power law continuum. Four absorption
features common to both plots are consistent with an outflow velocity in the range  6000-10000 km s$^{-1}$ when identified with the Lyman-$\alpha$ line of -
from left to right - Mg, Si, S and Ar. Deviations, seen particularly in the MOS data near 1.8 keV and 2.3 keV,  are attributed to  imperfect modelling
of the Si K- and gold M-absorption edges in the detector response and mirror reflectivity}      
\end{figure}

\subsection{The Fe K region}

Narrow absorption lines are also seen in the Fe K region of the EPIC spectra.  Correct modelling of the continuum is especially important in the neighbourhood of the Fe K lines as spectral curvature below
$\sim$6.4 keV will have an associated Fe-K absorption edge if attributed to reflection or partial covering. Cold reflection was shown in Po04 to dominate
the hard lowflux spectrum of \ngc\ in 2002, a conclusion consistent with the 2009 observation. It seems a reasonable assumption that a  similar underlying
component will contribute to the spectral curvature - and the Fe K emission line - in the present highflux data. 

Figure 7 shows a ratio plot of the EPIC pn highflux data to a continuum fitted at 2-10 keV with a power law, plus cold reflection fixed at the value
(R$\sim$1) determined from the lowflux data. Gaussian fitting finds the Fe K emission line can be modelled with narrow and broad components, with 4 possible
absorption lines detected to higher energy. Details of Gaussian fits to the Fe K spectral structure are listed in Table 1. Of particular interest here are
the absorption lines, where alternative conclusions can be drawn regarding outflow  velocities, depending on the line identifications. Though not
unambiguous, an appealing interpretation is where 2 velocity groups are found, with the lower value aligning with the higher velocities seen in the RGS
data. A still higher velocity, close to the value v$\sim$0.1c which may be a signature of high velocity winds (Tombesi \et\ 2010), is also
implied, though this depends on the significance of the 2 weaker absorption features observed at $\sim$7.5 and $\sim$7.7 keV. We return to the likelihood of
such a high velocity wind for \ngc\ in Section 6.2. 

\begin{figure}                                                          
\centering                                                              
\includegraphics[width=6.28cm, angle=270]{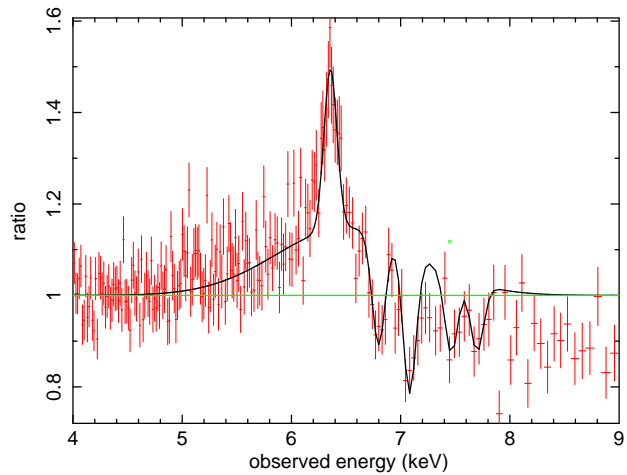}                                            
\caption                                                                 
{Gaussian fitting to spectral structure in the Fe K band of the highflux pn data, plotted against a power law plus reflection fit to  the continuum,
models the
Fe K emission line with broad narrow and broad components, while 4 possible absorption lines are detected to higher energy. Details are
listed in Table 1}      
\end{figure}

\begin{table}
\centering
\caption{Summary of Gaussian line fits (adjusted to the \ngc\ rest frame) to the Fe K spectral structure shown in figure 7}
\begin{tabular}{@{}lcccc@{}}
\hline
line & energy (keV) & width (eV)   & ident  & velocity (km s$^{-1}$ \\
\hline
1 & 6.38$\pm$0.01 & 60 (f) & Fe-K & - \\
2 & 6.4$\pm$0.1 & 380$\pm$63 & Fe-K &  -  \\
3 & 6.81$\pm$0.02 & 60 (f) & FeXXV &  -4900$\pm$750 \\
4 & 7.10$\pm$0.01 & 60 (f) & FeXXVI &  -5800$\pm$450 \\
5 & 7.49$\pm$0.03 & 60 (f) & FeXXV & -31500$\pm$2000 \\
6 & 7.72$\pm$0.03 & 60 (f) & FeXXVI & -29500$\pm$2000 \\
\hline
\end{tabular}
\end{table}

Summarising the EPIC absorption line data, we again emphasise that the deduced outflow velocities depend on the correct identification of less well resolved
absorption features (than for the RGS spectra). Identifying the absorption features in Mg, Si, S and Ar with the relevant Lyman-$\alpha$ line is the most
conservative, in the sense of yielding the lowest velocities, showing the intermediate velocity outflow seen in the RGS spectra  extends to a still higher
ionisation parameter, log$\xi$$\sim$3. Interpreting the Fe K absorption is more ambiguous and we retain both the intermediate and high velocity 
values in Table 2.

\begin{table*}
\centering
\caption{Absorption lines identified in the 2009 highflux \xmm\ spectrum of \ngc. The rest wavelength (energy) of each absorption line is in Angstroms (keV)
and measured line-of-sight outflow velocity components v1, v2, v3, v4, v5 are in km s$^{-1}$. Log$\xi$ (erg cm s$^{-1}$) is the ionisation parameter at which 
each ion stage
would have a maximum concentration in a photoionised gas illuminated by the AGN continuum flux. EPIC values are a mean of the pn and MOS fits for Mg,
Si, S and Ar while only pn values are shown for Fe. Only statistical errors are included}

\begin{tabular}{@{}lccccccc@{}}
\hline
Line i.d &  rest wavelength(energy) & v1 & v2 & v3 & v4 & v5 & log $\xi$ \\

\hline
CV 1s-3p & 34.97 & 250$\pm$60 & - & - &  - & - & 0\\
CVI Ly$\alpha$ & 33.74 & 490$\pm$35 & 4000$\pm$70 & 5900$\pm$90 & 8920$\pm$180 & - & 0.8\\
NVI 1s-2p & 28.79 & 325$\pm$40 & - & - & - & - & 0.45  \\
CVI Ly$\beta$ & 28.47 & 260$\pm$40 & - & - & - & - & 0.8 \\
NVI 1s-3p & 24.90 & 325$\pm$50 & - & - & - & - & 0.45  \\
NVII Ly$\alpha$ & 24.78 & 495$\pm$50 & 4000$\pm$80 & 5920$\pm$135 & 8600$\pm$120 & - & 1.1\\
OV 1s-2p & 22.37 & 310$\pm$60 & - & - & - & - & -0.25 \\
OVI 1s-2p & 22.01 & 440$\pm$75 & - & - & - & - & 0.2\\
OVII 1s-2p & 21.60 & 440$\pm$60 & 4080$\pm$140 & 5810$\pm$120 & - & - & 0.9 \\
OVI 1s-2p & 19.34 & 850$\pm$200 & - & - & - & - & 0.2\\
OVI 1s-2p & 19.18 & 660$\pm$200 & - & - & - & - & 0.2\\
OVIII Ly$\alpha$ & 18.97 & 780$\pm$70 & 4100$\pm$85 & 5600$\pm$60 & 9000$\pm$300 & - & 1.55  \\
OVIII Ly$\beta$ & 16.01 & 500$\pm$130 & 3700$\pm$130 & 6300$\pm$160 & - & - & 1.55 \\
NeIX 1s-3p & 11.55 & - & - & 5000$\pm$350 & - & - & 1.4 \\
NeX Ly$\alpha$  & 12.13 & 620$\pm$150 & - & - & 8230$\pm$130 & - & 2.0 \\
MgXII Ly$\alpha$ & (1.47) & - & - & -& 7100$\pm$1500  & - & 2.25 \\
SiXIV Ly$\alpha$ & (2.00) & - & - & -& 7100$\pm$2000 & - &  2.65 \\
SXVI Ly$\alpha$ & (2.62) & - & - & - & 7700$\pm$1600 & - & 2.9 \\
ArXVIII Ly$\alpha$ & (3.32) & - & - & - & 9700$\pm$1800 & - & 3.2 \\
FeXXV 1s-2p & (6.70) & - & - & 4900$\pm$750 & - &  31500$\pm$2000 & 3.3 \\
FeXXVI Ly$\alpha$ & (6.96) & - & - & 5800$\pm$450 & - & 29500$\pm$2000 & 3.9 \\

\hline
\end{tabular}
\end{table*}

\section{Modelling with XSTAR}

To further quantify the photoionised absorption in the complex outflow in \ngc\ we have modelled the highflux RGS spectrum with XSTAR. We again fitted  the
continuum with a power law, adding a negative Gaussian line to represent the broad UTA evident in figure 1 near 16 \AA. As in the Gaussian fitting, broad
emission features (line plus RRC) were added to the continuum.

A sequence of photoionised absorbers represented by grid 18 from the XSTAR library (Kallman \et\ 1996) was then added to model the absorption line spectrum.
Grid 18 includes a turbulent velocity  of 100 km s$^{-1}$, with an  ionising power law spectrum similar to that observed for \ngc\, irradiating a gas taken
initially to be of solar abundance. The ionisation parameter, column density and velocity (output as a modified redshift) are the primary free parameters of
each absorber. 

We began by modelling the RGS 1 data over the waveband 17-24 \AA, dominated by absorption lines of OIV, V, VI, VII and VIII and allowing a fit  largely
independent of relative abundances. A visually acceptable fit was obtained with 2 photoionised components expressing the low and intermediate velocities. The
fit was then extended to 36 \AA, to include the resonance lines of  NVII,NVI and CVI, and with both RGS1 and RGS2 data, and finally over the full waveband
10-36 \AA, covering the higher energy K-shell resonance transitions of Ne, and a potential complex of Fe-L lines. For the full band fits the abundances were
allowed to vary, although tied for the same element across the separate ionised components.

Five photoionised absorbers were required to obtain a satisfactory spectral fit over the whole waveband,  with a statistical improvement from
$\chi^{2}$/d.o.f. = 6788/4204 to  $\chi^{2}$/d.o.f. = 5543/4180. The abundances of the most important elements (relative to oxygen=1) were  C: 0.53$\pm$0.16, 
N:2.4$\pm$0.5, Ne: 2.6$\pm$2.0  and Fe:1.30$\pm$0.26.

The parameters of this multi-absorber fit are listed in Table 3 in the order in which they were added. Components 1 and 2 evidently represent 
the high velocity continuum absorption observed most strongly in OVIII and higher level ions.  

Component 3 represents the low velocity and lower ion stage continuum absorption, as well as the low velocity self-absorption in the broad line emission (figure
10).
Component 4 reflects the existence of low velocity absorption
across a wide range of ionisation parameter. We interpret that in a more detailed study of the broad line emission (Paper II) where the emission and strong
self-absorption arise from a limb-brightened shell, the velocities being limited by the high-inclination to the line of sight (figure 10).

Component 5 is intriguing in indicating a small redshift. We suggest that the red-shifted absorption, which importantly can also be seen in the
data, arises from the decellerating outflow on the far side of the AGN, the low ionisation parameter being consistent with an origin in the higher density
gas accumulating ahead of the contact discontinuity (see Discussion). 

Components 1-4 are added in figure 8 to a plot of outflow velocities obtained from the Gaussian line fitting, where each ion is located at the ionisation
parameter where it would have maximum abundance. We have not included the uncertainties on each XSTAR component as the fit should be considered as
representative of a continuous velocity and ionisation distribution. Nevertheless, as discussed in Section 6, we believe the near linear progression
from components 1 to 3 is an important demonstration of a mass conserved outflow.

A 5th asterisk in figure 8 represents the putative pre-shock high velocity flow, with velocity v$\sim$0.1c and ionisation parameter, adjusted to the current
sub-Eddington luminosity, of log$\xi$$\sim$3.7.  

\begin{table}
\centering
\caption{Parameters of the photoionised outflow fitted to the RGS data. Column density is in H atoms cm$^{-2}$ and
ionisation parameter in erg cm s$^{-1}$}

\begin{tabular}{@{}lccccc@{}}
\hline
Comp & log$\xi$ & N$_{H}$  & velocity (km s$^{-1}$ & $\Delta$$\chi^{2}$\\

\hline
1 & 2.97$\pm$0.06 & 1.4$\pm$0.2$\times10^{22}$ & -5880$\pm$60 & 484 \\
2 & 2.52$\pm$0.07 & 1.85$\pm$0.4$\times10^{21}$ & -3850$\pm$60 & 178 \\
3 & 1.43$\pm$0.22 & 1.1$\pm$0.5$\times10^{20}$ & -530$\pm$90 & 394 \\
4 & 2.77$\pm$0.15 & 2.9$\pm$1.4$\times10^{21}$ & -400$\pm$50 &  64 \\
5 & 0.32$\pm$0.24 & 1.0$\pm$0.2$\times10^{20}$ & +120$\pm$60 & 123 \\
\hline
\end{tabular}
\end{table}

\section{Discussion}

A striking feature of the absorption spectrum in \ngc\ is the very wide range of velocities and ion stages observed. A second feature is the correlation of
velocity and ionisation parameter in the continuum absorption. 

Figure 9 visualises that correlation, with contrasting velocity profiles 
of the opacity in OVIII and OVII, for the high flux revs 1722 and 1724. The upper plot, centred at the rest wavelength of OVIII Lyman-$\alpha$,  shows the onset of
absorption at $\sim$7500 km s$^{-1}$, with the opacity increasing to $\sim$6000--5000 km s$^{-1}$, whereafter it decreases again to disappear at 
$\sim$3500 km s$^{-1}$, re-emerging strongly below $\sim$1500 km s$^{-1}$.  The corresponding velocity profile in the OVII 1s-2p
resonance line shows only weak high velocity opacity at $\sim$4500--3000 km s$^{-1}$. In contrast the low
velocity absorption is notably stronger in the lower energy ion

Figure 8 brings together the velocity
data from Gaussian fitting to the RGS and EPIC absorption spectra, plotting each well-defined velocity against the optimum  ionisation parameter for the parent ion. The
strongest absorption components at each velocity generally lie to the right hand - higher ionisation - side of the plot, consistent with a broad correlation
of velocity and ionisation parameter, while the highest velocities are only detected in the higher energy parent ions, and the low velocity group are seen
most strongly - or only - in low ionisation matter.

The correlation of velocity and ionisation parameter is further illustrated by the results from XSTAR modelling of the highflux absorption spectrum, represented by
asterisks in figure 8, with components  1, 2 and 3 following  a clear linear trend in velocity and ionisation parameter .
Figure 8 also indicates that part of the low velocity absorption does not follow the general trend with ionisation, with XSTAR component 4
supporting the evidence from Gaussian fitting that low velocity absorption exists at all ionisation stages covered by the  RGS data. Figure 10 shows
schematically how separate absorption spectra are associated with the direct continuum and with self absorption in the post shcck gas. More detailed
consideration of a low velocity absorption structure, visible across all ion stages in the RGS data (and represented by components 3, 4 and 5 in the XSTAR
modelling), is deferred to Paper II where it is interpreted as self-absorption in a limb-brightened shell.

Meanwhile, an overview of figure 8 suggests 3 broad velocity regimes in the outflow in line-of-sight to \ngc, including the higher velocity component in  the
Fe K  absorption.  On this overview, the individual velocity components picked out by Gaussian fitting at $\sim$4000 km s$^{-1}$, $\sim$6000 km s$^{-1}$ and 
$\sim$8500 km s$^{-1}$, but appearing more like a broad trough in the strongest absorption, in OVIII Lyman-$\alpha$ (figure 9), may represent density variations  within
a continuous flow. Further evidence for such density variations, which might represent a residual shell-structure  linked to the intermittent nature of the
initial fast wind, can be seen in velocity profiles taken at different flux levels for different ions.

\begin{figure}                                                                                                                                            
\centering                                                              
\includegraphics[width=6.1cm, angle=270]{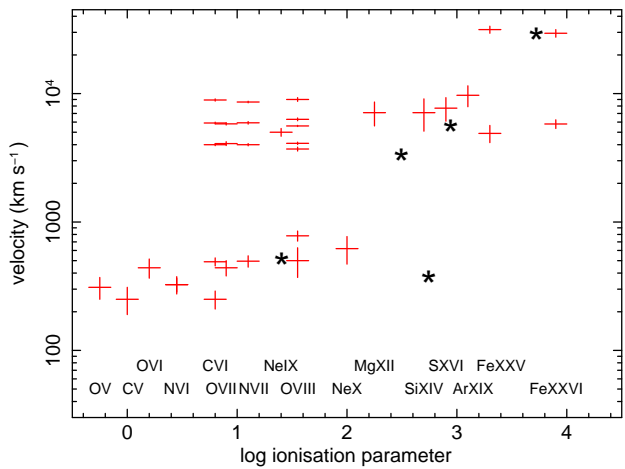}                                                                                                        
\caption                                                                
{Outflow velocities derived from the Gaussian fitting plotted against the optimum ionisation parameter for each parent ion stage. Also shown by asterisks are the
parameters of the 4 photoionised absorbers derived from XSTAR modelling of the RGS absorption spectra, together with a velocity/ high ionisation point to
represent the putative pre-shock wind}       
\end{figure} 

In the remainder of this paper we refer to the low ($\leq$1000 km s$^{-1}$), intermediate ($\sim$3000-9000 km s$^{-1}$) and high velocity  ($\sim$30000 km
s$^{-1}$) regions indicated in figure 8 and explore the possibility that they represent different stages in a shocked outflow. The implications of such an
interpretation are doubly significant. The concept of a slowing and cooling/recombining ionised outflow is in contrast to most current ideas for the radial
acceleration of AGN winds, while efficient post-shock cooling would mean that the mechanical energy in a fast outflow may not be the primary mechanism for
AGN feedback.

\begin{figure}
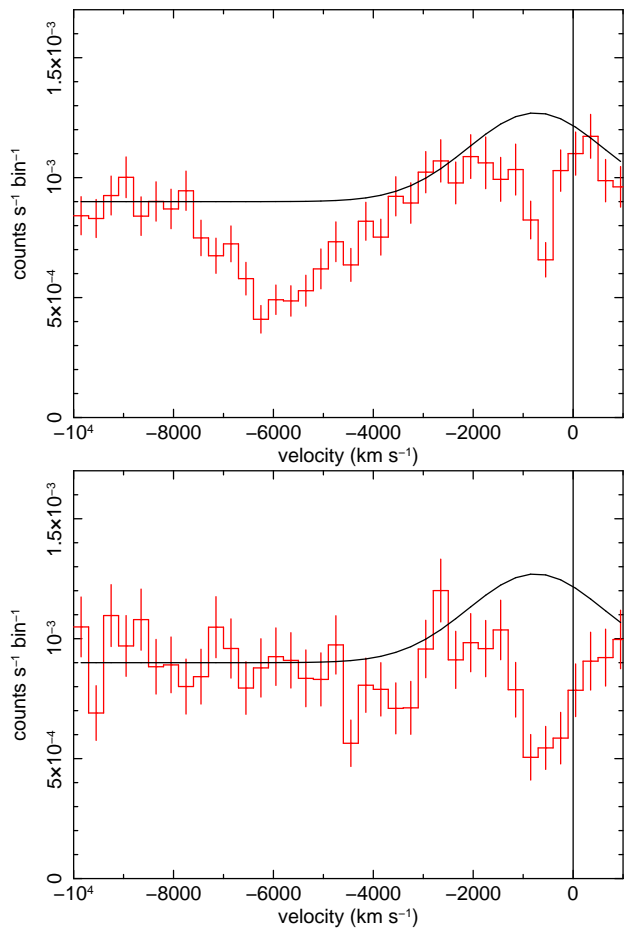
                                                          
\centering                                                              
\includegraphics[width=6.1cm, angle=270]{newfig9.ps}                                                                                   
\centering                                                              
\includegraphics[width=6.1cm, angle=270]{newfig9a.ps}                                                                                                                                                                      
\caption                                                                
{Contrasting velocity profiles of the opacity in OVIII and OVII, for the high flux revs 1722 and 1724, illustrate the strong correlation of velocity and 
ionisation parameter consistent with a decellerating post-shock flow. The upper plot, centred at the rest wavelength of OVIII Lyman-$\alpha$,  shows the onset of
absorption at $\sim$7500 km s$^{-1}$, with opacity increasing to $\sim$6000--5000 km s$^{-1}$, whereafter it decreases again to disappear at 
$\sim$3500 km s$^{-1}$, re-emerging strongly below $\sim$1500 km s$^{-1}$.  The corresponding velocity profile in the OVII 1s-2p
resonance line shows significant high velocity opacity only at $\sim$4500--3000 km s$^{-1}$. In contrast the low
velocity absorption is notably stronger in the lower energy ion}       
\end{figure}

\subsection{Outflows and feedback in AGN}

Comparison with previous observations of \ngc\ by \xmm\ and \chandra\ suggest that the low velocity absorption is persistent, while that at higher velocities
is probably variable, at least on a timescale of years. The caveat on variability is that most of the earlier observations were of lower sensitivity, while
the relatively long 2001 \xmm\ observation does - on a re-examination - show evidence of blue shifted absorption corresponding to an outflow in the range
$\sim$4000-6000 km s$^{-1}$.  The new \xmm\ spectra are most closely approached by the \chandra\ LETG spectra reported by Steenbrugge (2009), who find
outflow velocities of $\sim$200, $\sim$600 and $\sim$4600 km s$^{-1}$, modelled by 4 ionised absorbers with ionisation parameters ranging from
log$\xi$$\sim$0--3.    

In discussing the 2001 RGS spectrum of \ngc\ Po04 noted that, with the simple assumption of conservation of mass in a  radial outflow, an extended region of
slow moving, low ionisation gas might be a continuation of the high velocity, high ionisation flow seen in absorption in the Fe K band. On this picture much
of the mechanical energy in the initial outflow would  have been lost before reaching the lower ionisation stage, and Po04 speculated that this might be due
to internal shocks occurring in the high velocity gas. 

We now take up that idea again but in the context of a shock interaction with slower moving matter, either the local ISM or previous ejecta. To justify that
approach we note that the broad correlation of velocity and ionisation state in the absorption spectra for \ngc\ is a clear signature of a cooling,
decellerating and recombining outflow. 

The mass rate in a radial outflow of velocity v, and particle density n at radius r is $\mo = 4\pi bnr^2.v.m_{p}$, where b is the fractional collimation
angle and $m_{p}$ is the proton mass.  Mass conservation in the flow requires the product n.r$^{2}$.v to be constant. As n.r$^{2}$ = L$_{ion}$/$\xi$, if the
ionising radiation is unchanged (or changes very little), for example over a distance small compared with r,  then conservation of mass in a radial outflow
yields a linear correlation of velocity and ionisation parameter. We would expect that to be the case in a post-shock cooling shell. The important
implication of a decellerating radial outflow is that the mechanical energy in the flow would be substantially reduced as the flow
is slowed.

King (2010) has recently examined a relevant scenario, where a highly ionised, high velocity wind drives into the interstellar medium,  losing much of its
energy by efficient cooling of the shocked gas. 
Such a scenario would have major implications for studies of AGN feedback based on X-ray absorption spectra.  Until now, X-ray observations have been used
in attempts to show that  fast, ionised outflows can provide the link  between the growth of a SMBH and its host galaxy, by the integrated mechanical energy
in the fast flow (eg Pounds and Reeves 2009). However, as recently pointed out (eg King 2010),  if such an energetic wind persists while the black hole
doubles its mass (the Salpeter time), the coupling of wind energy to galactic baryons may have to be inefficient to allow massive bulges to grow to the
values observed, effective feedback instead being enabled by the total
momentum of the flow, an alternative that has been shown to yield the observed M - $\sigma$ relationship for nearby active galaxies (King 2003, 2005).

In the following section the new \xmm\ observations of \ngc\ are examined in the context of a shocked wind,
where the intermediate velocity/intermediate ionisation outflow corresponds to the immediate post-shock gas and the low velocity/low ionisation absorption
to matter building up ahead of the contact discontinuity. 

\subsection{Comparing the \ngc\ data with a shocked wind model}

In the King (2010) shocked wind model a high velocity ionised outflow collides with the ISM of the host galaxy, resulting in a strong shock. The gas density
increases by a factor $\sim 4$ at the shock front, and the velocity drops by the same factor. Beyond this (reverse, adiabatic) shock, the flow is further
compressed in a relatively thin, cooling region, while the velocity slows to low values. Strong Compton cooling by the AGN radiation implies a fairly rapid
transition  between the  immediate post-shock regime and the much slower and compressed state near the contact discontinuity.  Beyond the contact
discontinuity a further low velocity, low ionisation component will build up as the interstellar medium is swept up by an outer (forward) shock. 

Comparison of that scenario with the new \ngc\ data assumes an intermittent  highly ionised wind, with typical values of v$\sim$0.1c and
log$\xi$$\sim$4 (Tombesi \et\ 2010), has collided with the ISM or previous slower moving ejecta, with the density increase rendering the immediate
post-shock gas, v$\sim$0.025c ($\sim$7500 km s$^{-1}$), visible as resonance absorption at log$\xi$$\leq$3. 

The shocked wind cools and slows after passing through the inner shock, yielding  the broad absorption trough seen in the velocity plot for OVIII
Lyman-$\alpha$ line (figure 9). We speculate that the observed gap between the intermediate and low velocity absorbing material arises from the falling
column density of OVIII, re-emerging at low velocities as the column density builds up ahead of the contact discontinuity.   

It is interesting to consider how the observations of \ngc\ may be used to estimate the  parameters of an earlier, perhaps intermittent, Eddington episode
whose effects we may now be observing. While direct evidence for a high velocity wind is marginal in the 2009 data, as it was in the 2001 \xmm\ observation,
Tombesi \et\ report the detection of an outflow at $\sim$0.13c in the 2002 observation. Taken together, 
those reports indicate the present fast outflow in
\ngc\ is intermittent, and we note  that with the unusually high column density required to detect a blue-shifted Fe K absorption line in a low redshift source
such as \ngc,  radial expansion would rapidly render a transient outflow undetectable with EPIC. 

In what follows, we explore the effects of such an intermittent  fast  outflow as it interacts with the ISM or slower moving ejecta, as an explanation of
the velocity- and ionisation-structured outflow phases observed in the present observation of \ngc.

\subsection{The intermediate ionisation/intermediate velocity gas}

We identify the immediate post-shock outflow in \ngc\ with the onset of substantial opacity at a velocity  v$\sim$$7\times10^{8}$ cm s$^{-1}$ (figure 9, top
panel).
XSTAR modelling (Table 3) finds an ionisation parameter log$\xi$$\sim$3 and column density of  N$_{H}$$\sim$$1.4\times 10^{22}$ cm$^{-2}$ to represent this
flow phase. Both velocity and ionisation parameter are consistent with the factor $\sim$4 change from the putative 0.1c wind  expected across a strong
shock.    

The incident ionising luminosity of $\sim$$8\times10^{41}$ erg s$^{-1}$, for an average-flux EPIC spectrum (rev 1729), together with the fitted ionisation
parameter and measured velocity, give nr$^{2}$v $\sim$$6\times 10^{47}$ s$^{-1}$ for the post-shock flow.  Assuming an angular collimation b=0.3, the
post-shock flow mass rate is then $\mo$ $\sim$$4\times 10^{24}$ gm s$^{-1}$ ($\sim$0.06 $\msun$ yr$^{-1}$). Interestingly, that rate is close to the Eddington
accretion rate for \ngc\ assuming an accretion efficiency of 0.1.

The corresponding momentum rate of the intermediate velocity gas is then $\sim$$3\times 10^{33}$ (cgs), with mechanical energy (0.5$\mo.v^{2}$) of
$\sim$$10^{42}$ erg s$^{-1}$.  That mechanical energy rate is $\sim$0.3$\%$ of the Eddington luminosity for a black hole mass of   $\sim$$1.7\times
10^{6}$\Msun (Denney \et\ 2009), a factor $\sim$30 less than the value v/c.L$_{Edd}$   predicted by a simple continuum driving model (King and Pounds 2003),
roughly consistent with the expected velocity-linked loss at the strong shock.  

As the post-shock gas cools we can identify second and third XSTAR components on the velocity/ionisation plot (figure 8), with v$\sim$3850 km s$^{-1}$ and
log$\xi$$\sim$2.55, and v$\sim$550 km s$^{-1}$ and log$\xi$$\sim$1.56, following reasonably closely the expectation of a linear correlation of velocity and
ionisation parameter in a post-shock flow. 

The detection of several strong RRC and broad resonance line emission is indicative of a strongly recombining stage in the flow,  
and the observed OVII RRC flux of $\sim$$4\times 10^{-5}$ photons cm$^{-2}$s$^{-1}$ provides a measure of the intermediate velocity flow. A
temperature of $\sim$5 eV, from the width of the RRC, implies a recombination rate for OVIII of $\sim$$10^{-11}$~cm$^{3}$~s$^{-1}$ (Verner and Ferland
1996). With that value and assuming 30 percent of recombinations from the majority OVIII ion direct to the ground  state, we 
deduce an emission measure from the OVII RRC flux of order $\sim$$2\times10^{63}$cm$^{-3}$, for a Tully-Fisher distance  of 15.2 Mpc. 
We note, in passing, that this could be a minimum measure as any much higher temperature RRC component would be difficult to resolve.

Importantly, the particle density in the intermediate flow region is constrained by evidence of a narrowing of the broad emission component in OVIII 
in rev 1739, indicating a change in the ionisation state following a 3-4 day interval of unusually low  continuum flux level. This is discussed in more detail in
PaperII. In the
following estimates of the flow properties we assume a recombination timescale for OVIII of $\sim$4 days, corresponding to  a particle density
n$\sim$$5\times 10^{5}$cm$^{-3}$ for the intermediate velocity flow.

The OVII RRC emission measure then corresponds to an emitting volume of $\sim$$8\times10^{51}$cm$^{3}$ and - taking the column density of component 2 in
the XSTAR modelling as a measure of the intermediate velocity absorber - gives a radial thickness of  order $\Delta$r$\sim$$4\times10^{15}$ cm. 

At a mid-phase velocity of $\sim$4000 km s$^{-1}$, the shocked gas would traverse this cooling region in a time t$\sim$$10^{7}$ s. That would suggest the intermediate velocity absorption could exist in the absence of a fast outflow for a similar timescale, while the considerably
shorter recombination timescale could explain structure in the intermediate velocity flow, in turn perhaps reflecting the intermittent nature of the fast wind.

From the above estimates of particle density, shell thickness and emission volume (acknowledging that these are only a crude measure across a likely  strong
radial gradient), we find a shell radius  r$\sim$$7\times10^{17}$ cm, for b=0.3. 

We now compare the above parameters of the intermediate velocity/ionisation post-shock gas with values for the low ionisation/low velocity gas accumulating 
ahead of the contact discontinuity, to provide an order-of-magnitude estimate of the duration and history of the putative  Eddington accretion episode in
\ngc. 

\begin{figure}                                                                                                                                            
\centering                                                              
\includegraphics[width=7.7cm, angle=0]{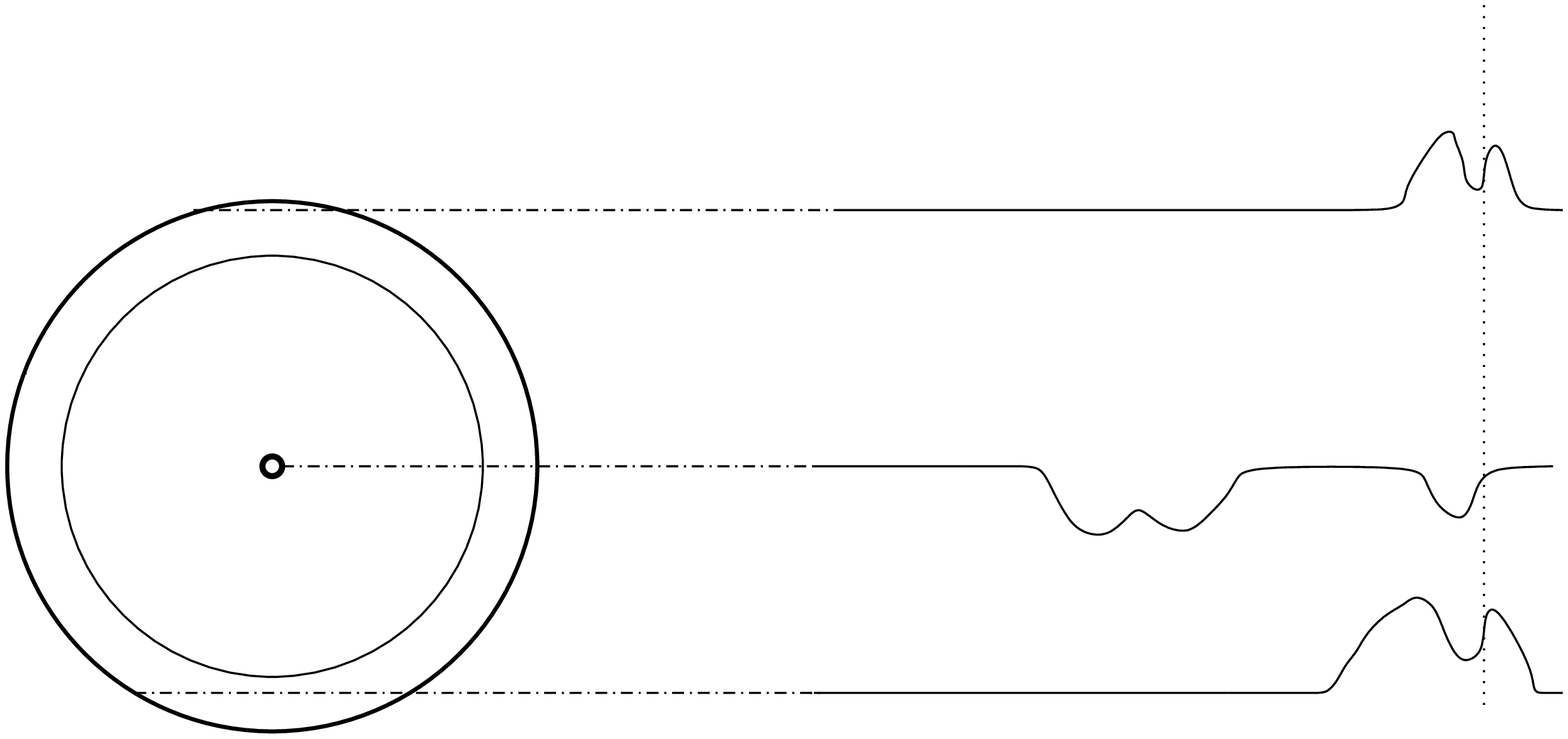}                                                                                                        
\caption                                                                
{Sketch showing the origin of separate absorption spectra, in the continuum by line-of-sight to the AGN and by self-absorption in the soft 
X-ray emission from a limb brightened
shell}       
\end{figure}

\subsection{Constraining the low ionisation/low velocity gas}

We take component 3 from the XSTAR photoionisation modelling to represent the low ionisation/low velocity gas accumulating ahead of the contact 
discontinuity (CD) and seen in continuum absorption in the highflux spectrum. That component has an ionisation  parameter of log$\xi$$\sim$1.43, absorbing
column density of $\sim$$10^{20}$cm$^{-2}$, and outflow velocity of $\sim$530 km s$^{-1}$. With a mean ionising luminosity ($\ga$l keV) of $8\times
10^{41}$ erg s$^{-1}$, we obtain n.r$^{2}$$\sim$$3\times 10^{40}$ for the low velocity/low ionisation flow gas. With the particle density scaling from the
velocity difference - and reflected  in the lower ionisation parameter, we assume  n$\sim$$5\times 10^{6}$cm$^{-3}$, giving a radial depth of the pre-CD
shell $\delta$r $\sim$ $2\times 10^{13}$ cm$^{-2}$.

The higher assumed density of the pre-CD shell would appear to conflict with the persistence of the low velocity absorption, particular in a higher  level
line such as OVIII Lyman-$alpha$. However, as we report in Paper II, self-absorption of the broad emission line can fully account for the low
velocity opacity when the continuum is weak.

As the radius of this low ionisation shell must exceed the estimated thickness of the intermediate flow region, we note that light travel time
delays would further limit variability in the low velocity opacity, consistent with the lower limit of r$\ga$$9\times10^{17}$ cm obtained  by Steenbrugge
\et\ (2009) from a lack of variability in the low velocity component in the \chandra\ absorption spectrum. We assume r$\sim$$10^{18}$ cm below.

With that geometry the mass of accumulated low  ionisation gas, seen in absorption, is $\sim$$7\times 10^{32}$ gm ($\sim$0.3\Msun). Comparison with the 
mass rate in the post shock flow indicates an accumulation time of $\sim$6 years. For a velocity of 530 km s$^{-1}$, the mechanical energy in this
component of the low velocity gas is then $\sim$$10^{48}$ergs, indicating that   $\sim$99.5\% of the  mechanical energy has been lost in the
post-shock cooling.   

While the total flow momentum will be conserved through the shock, we expect a major fraction of the initial ram pressure is converted to gas pressure at
the contact  discontinuity.  Integrating the immediate post-shock outflow momentum over 6 years totals $\sim$$10^{41}$ (cgs). In comparison the accumulated
ram pressure in the post-shock, low ionisation gas is $\sim$$3.7\times10^{40}$ (cgs), indicating  $\sim$63\% of the immediate post-shock  momentum has been
translated into pressure ahead of the contact discontinuity.

\subsection{Evidence for radiation from the shocked gas}

If the shocked outflow scenario does apply to \ngc\ the question arises as to whether there is direct evidence for radiation from the post-shock cooling?

The high temperatures in the shock require strong cooling which is likely to be dominated by Compton scattering of the AGN's radiation field.  King (2010)
finds that this  typically will have a Compton temperature $T_c\sim 10^7$~K, compared with the much higher adiabatic shock temperature of  $m_pv^2/k$
$\sim$$10^{11}$~K. 

In assessing whether there is any direct evidence of such cooling radiation, we recall that  Uttley \et\ (2003) identified a quasi-constant soft X-ray
component, modelled as a $\Gamma$$\sim$3 power law, in a \chandra\ TOO observation made during a 6-week low flux state of \ngc. The overall X-ray spectrum
in the present low flux rev1739 data is very similar to the \chandra\ observation, and also to that during the \xmm\ observation  in 2002 which followed a
20-day low flux period. Furthermore, regular monitoring with Swift  during the present \xmm\ campaign indicated that rev1739 also  followed a low flux state
lasting for several days. We therefore take the rev1739 spectrum of \ngc\ to be typical of a possible base level spectrum. 

Figure 11 shows the pn data from rev1739 fitted with two continuum components.  Above the sharp spectral break at $\sim$1-2 keV the spectrum is
parameterised by a hard power law ($\Gamma$$\sim$0.8), while the steep soft X-ray component can be modelled by either a much softer power law
($\Gamma$$\sim$3.5), or a Comptonised spectrum (as shown in the figure) with kT$\sim$0.3 keV and optical depth $\tau$$\sim$0.3. A strong Gaussian emission
line (equivalent width 250$\pm$30 eV) at 6.38$\pm$0.01 keV is consistent with Fe K fluorescence from the hard spectral component being reflection dominated,
as reported in Po04. 

The short-term flux  variability seen throughout the rest of the 2009 \xmm\ observation of \ngc\ is notably absent in rev 1739 (Vaughan \et\ 2010), 
supporting the view that the soft X-ray component in figure 11 does indeed comprise a quasi-constant low flux emission, with the lack of
variability indicating its origin in an extended region. The post-shock cooling shell could fit that description and it is notable that the present RGS
spectra show that absorption is limited to self-absorption in the broad emission lines at the lowest continuum fluxes. The luminosity in the
Compton component of figure 11 ($\ga$0.2 keV) is $\sim$$3\times10^{41}$ erg s$^{-1}$, comparable to the mechanical energy lost in traversing the
post-shock region.  

We have suggested that the broad emission lines  and recombination continua (RRC) observed  in the RGS spectrum arise from additional 2-body cooling of the
post-shock outflow. Fitting in XSPEC finds fluxes of $\sim$$1.5\times10^{-4}$ photons cm$^{-1}$ s$^{-1}$ and  $\sim$$2\times10^{-5}$ photons cm$^{-1}$
s$^{-1}$, respectively, for the broad emission lines of OVII and OVIII Lyman-$\alpha$. Addition of the corresponding broad emission lines of Ne, N and C
yields a total soft X-ray flux of  $\ga$$10^{-3}$ photons cm$^{-1}$ s$^{-1}$, and a luminosity $\ga$$2\times10^{40}$ erg s$^{-1}$.    Strong RRC of O, N and
C and emission from Fe-L lines increase the total observed recombination cooling to $\sim$$5\times10^{40}$ erg s$^{-1}$, making a significant addition to
the cooling in the later stages of the post-shock flow.  

We  conclude that identifying a quasi-constant soft continuum and soft X-ray emission features in \ngc\ with the cooling of the shock outflow is reasonable
on energetic grounds.

\begin{figure}                                                          
\centering                                                                                                                                    
\includegraphics[width=6.1cm, angle=270]{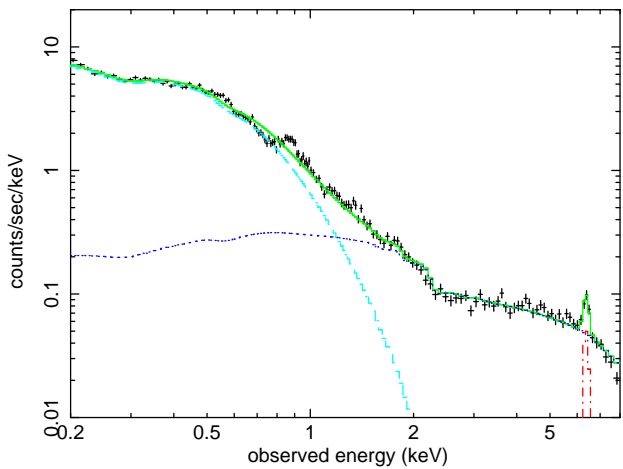}                      
\caption                                                                
{Parametric model fit to the lowflux rev1739 pn data. The hard power law and soft Comptonised continuum components are
shown as dotted and dashed lines}       
\end{figure}

\subsection{Relating \ngc\ to other AGN with powerful ionised winds}

Evidence has been growing for ultra-fast winds of highly ionised matter in a number of AGN, carrying mechanical energy up to 10$\%$ of L$_{bol}$ ( Pounds
and Reeves 2009). The direct determination of a large-angle flow in the QSO PG1211+143 was important in confirming that the v$\sim$0.13c wind was
energetically significant in terms of AGN feedback. Indirect support for such fast outflows to be typically of wide angle has recently been obtained  in a
survey of bright AGN by Tombesi \et\ (2010), who find some 30\% of their sample show evidence for an ionised wind of v$\sim$0.1c.  As noted above, the
potential importance of such powerful winds lies in providing a feedback mechanism linking the growth (and termination of growth) of supermassive black
holes in AGN with  that of their host galaxy. This remains true if the impact is actually delivered by a momentum-driven thrust, as would be the case if
the initial ionised wind lost much of its mechanical energy in shocks before reaching the star-forming region. The analysis outlined above suggests that may
be the case for the bright, nearby Seyfert galaxy \ngc.

The question of how a highly ionised gas is accelerated to such high velocities is clearly important. Continuum driving for AGN accreting at a
super-Eddington rate was proposed by King and Pounds (2003), and appears readily applicable to  PG1211+143. However, in the case of many of the Tombesi \et\
sample, including \ngc, current mass estimates suggest they are accreting more typically at only $\sim$10-20 \% of L$_{Edd}$.  

If the Tombesi \et\ findings are confirmed, with highly ionised outflows at v$\sim$0.1c being relatively common in bright nearby galaxies, then perhaps
Eddington or mildly super-Eddington accretion is also more common than generally believed. A case for AGN black hole masses being over-estimated has
recently been argued by King (2010a). Furthermore, if the ejection of fast outflows is intermittent, as suggested here for \ngc, then only where such a 
wind is current or was launched very recently will it retain a line-of-sight column close to the theoretical value for continuum driving of
N$_{H}$$\sim$$10^{24}$ cm$^{-2}$.

\subsection{Is the forbidden line emission dominated by swept up ISM?}

Strong and narrow emission lines in the RGS data, of similar strength in both low and high flux spectra, arise from the `forbidden' transitions in the 1s-2p
triplets of He-like NVI, OVII and NeIX. In each case we find the line is at best only marginally resolved, with the strongest OVII line  indicating a 
FWHM$\leq$250 km s$^{-1}$. When adjusted for the known redshift of \ngc\ the OVII forbidden line also has a very low outflow velocity of -125$\pm$40 km
s$^{-1}$. We note that the velocity width is consistent with the [OIII] line width in the NLR, of 210-330  km s$^{-1}$ FWHM (De Robertis and Osterbrock
1984). 

In P04  constraints on the low ionisation/low velocity gas were obtained from the OVII emission line flux by noting that the 2002 November \xmm\ observation
took  place some 20 days after the source entered an extended low flux state, while the emission line strength of the OVII forbidden line was essentially
the same as when \ngc\ was much brighter in 2001 May. This was taken to indicate a recombination time $\ga$$2\times10^{6}$s, and a plasma density 
n$\leq$$8\times 10^{4}$cm$^{-3}$. We now note the forbidden line fluxes in the 2009 RGS data are consistent with those measured 7 years earlier. If that is
a real  measure of  lack of variability it would indicate a still lower density n$\leq$$10^{3}$cm$^{-3}$, or a physical extent of the forbidden line
emission region larger than the present estimate for a post shock region of radius $\sim$0.3 pc.

The question then arises, does a substantial fraction of the forbidden line emission come from the slower moving swept-up ISM ahead of the outer shock? This 
possibility is raised by noting that the resonance absorption line of OVII, identified in the context of a shocked wind with matter ahead at the contact 
discontinuity, has a lowest velocity component of 440$\pm$60 km s$^{-1}$, well separated from that of the observed OVII forbidden line. Absorption in OV and
- less unambiguously - in OIV also have lower velocities and lie well to the low ionisation side of the linear correlation with velocity that fits the main
outflow stages in figure 8. 

Estimating the baryon mass of the swept-up ISM from the present data depends both on the assumed density and also - quite strongly - on the relevant
ionisation parameter. While the optimum ionisation parameter for OVII is log$\xi$$\sim$0.9, the very low velocity would associate it more with absorption
from OIV-VI and log$\xi$$\sim$0, where OVII would represent only $\sim$3 \% of total oxygen. 

With the measured OVII forbidden line flux of $\sim$$1.5\times10^{-4}$ photons cm$^{-1}$ s$^{-1}$, a distance to \ngc\ of 15.2 Mpc, 3\% of oxygen  in the
form of OVII, and a recombination rate at kT $\sim$3 eV of $2\times 10^{-11}$~cm$^{3}$~s$^{-1}$ (Verner and Ferland 1996) we find an emission measure of 
$\sim$$10^{64}$cm$^{-3}$.


For a particle density n$\sim$$10^{3}$cm$^{-3}$ that emission measure would correspond to a total swept-up mass  of $\sim$7500 \Msun.  Assuming an ISM
density close to the virial equilibrium value (i.e. the isothermal sphere value for $\sigma$ = 88 km $s^{-1}$ in \ngc), the swept up mass within 
r$\sim$$10^{18}$ cm, for a gas fraction of 0.15, would be $\sim$$3.5\times 10^{38}$gm  ($1.7\times 10^{5}$ \Msun. The implication is that previous Eddington
episodes in \ngc\ have substantially reduced the gas density in the  inner core of the galaxy.


\section{Summary}

Extended new \xmm\ observations of the Seyfert 1 galaxy \ngc\ have shown a rich X-ray absorption line spectrum from 16 K-shell ions from C to Fe. The
increased sensitivity and energy span of these observations reveal a velocity-structured outflow ranging over at least two orders of magnitude in velocity
and ionisation parameter. We find a strong correlation of velocity and ionisation parameter, with both parameters being structured in at least 3 regimes.

We assess the observations in terms of a mass-conserved outflow, where a highly ionised, high velocity wind  is slowed on impact with the local 
interstellar medium or slower moving ejecta, resulting in a strong shock.  While most of the initial mechanical energy is lost in the shock, the total 
outflow momentum is maintained.

On this picture, we associate the intermediate velocity components, observed at $\sim$3000-7500 km s$^{-1}$ and with the highest opacity at  intermediate
ionisation parameters, with the rapidly cooling shell (or shells) of post-shock gas, while a well-separated continuum component at $\sim$400-700 km s$^{-1}$, in a
lower ionisation state, is attributed to the gas accumulating ahead of the contact discontinuity. We speculate that a still lower velocity component
may be in the ISM swept up by the forward shock.

Compton cooling of the shocked gas by the AGN radiation field is predicted to produce a soft continuum emission containing much of lost outflow kinetic
energy and we propose that part of this might be observed as a residual, quasi-constant soft X-ray component  ($\Gamma$$\sim$3.5)  in the lowest flux
spectra of \ngc. 

The broad line emission and strong RRC observed from several of the most abundant ions in the RGS spectrum provide additional cooling as the post-shock flow
recombines. Self-absorption arising in the higher line-of-sight column density around the circumference of a thin shell of post-shock gas could explain the
persistence of low velocity absorption, across a wide range of ions, even when the continuum is faint.

If a non-varying OVII forbidden line is a measure of the swept-up ISM ahead of the forward shock, the estimated mass of $\sim$7500\Msun\ represents only
$\sim$4\% of that for a virial gas density, suggesting that previous Eddington episodes in  \ngc\ have substantially reduced the gas density in the  central
regions of the galaxy.

\section*{ Acknowledgements } 
The results reported here are based on observations obtained with \xmm, an ESA science mission with
instruments and contributions directly funded by ESA Member States and the USA (NASA). The authors wish
to thank Andrew King for many stimulating discussions, the anonymous referee for helpful suggestions in improving the structure of the paper, 
and the SOC and SSC teams for organising the \xmm\ observations and initial data reduction.

\end{document}